\newif\ifIEEE
\IEEEtrue
\IEEEfalse

\ifIEEE
\newenvironment{oproof}{\begin{IEEEproof}}{\end{IEEEproof}}
\newenvironment{omultline}{\begin{multline}}{\end{multline}}
\newenvironment{omultline*}{\begin{multline*}}{\end{multline*}}
\newcommand{\onl}{\\}  
\else
\newenvironment{oproof}{\begin{proof}}{\end{proof}}
\newenvironment{omultline}{\begin{equation}}{\end{equation}}
\newenvironment{omultline*}{\begin{equation*}}{\end{equation*}}
\newcommand{\onl}{}  
\fi

\ifIEEE
\documentclass[10pt,final,conference,leqno]{IEEEtran}
\else
\documentclass[a4paper,12pt,notitlepage]{scrartcl}
\fi

\usepackage[utf8]{inputenc}
\usepackage{amsmath}
\usepackage{amsthm}
\usepackage{amsfonts}
\usepackage{amssymb}

\usepackage[x11names,rgb]{xcolor}
\usepackage{tikz}
\usetikzlibrary{arrows}
\usetikzlibrary{shapes}
\usetikzlibrary{backgrounds}

\usepackage{url}
\usepackage{color}

\usepackage{verbatim}

\usepackage{hyperref}

\newtheorem{lemma}{Lemma}
\newtheorem{prop}[lemma]{Proposition}
\newtheorem{thm}[lemma]{Theorem}
\newtheorem*{thm*}{Theorem}

\theoremstyle{definition}
\newtheorem{defn}[lemma]{Definition}
\theoremstyle{remark}
\newtheorem{rem}[lemma]{Remark}
\newtheorem{ex}[lemma]{Example}

\newcommand{\Rb}{\mathbb{R}}

\newcommand{\Acal}{\mathcal{A}}
\newcommand{\Xcal}{\mathcal{X}}
\newcommand{\Ycal}{\mathcal{Y}}
\newcommand{\Zcal}{\mathcal{Z}}

\newcommand{\ol}{\overline}

\newcommand{\ones}{\mathbf{1}}
\newcommand{\zeros}{\mathbf{0}}

\newcommand{\Dia}{\ol{\zeros\ones}}

\newcommand{\uge}{\sqsupseteq}  
\newcommand{\mge}{\sqsupseteq'}  
\newcommand{\zge}{\succeq}  

\author{Nils Bertschinger, Johannes Rauh  \\
\small \{bertschinger,jrauh\}@mis.mpg.de \\
\small Max Planck Institute for Mathematics in the Sciences, \\
\small Leipzig, Germany}
\title{The Blackwell relation defines no lattice}
\date{October 17, 2013}

\begin{document}
\maketitle

\begin{abstract}
  Blackwell's theorem shows the equivalence of two preorders on the
  set of information channels. Here, we restate, and slightly
  generalize, his result in terms of random variables. Furthermore, we
  prove that the corresponding partial order is not a lattice; that
  is, least upper bounds and greatest lower bounds do not exist.
\end{abstract}

\section{Introduction}
\label{sec:introduction}

In 1953 Blackwell showed the equivalence of two preorders on the set
of information channels
\cite{Blackwell53:Equivalent_comparisons_of_experiments}. Given two
information channels $\mu, \kappa$ with the same source $X$, Blackwell
considers the maximal expected reward a rational agent can obtain when
decisions are based on the output of $\kappa$ or $\mu$
respectively. Now, $\kappa$ is defined to be {\em more informative}
than $\mu$ if, in any decision problem, the agent can perform  better by
using $\kappa$ instead of $\mu$. Blackwell's theorem provides an
equivalent characterization in terms of an algebraic
relation between the information channels: $\kappa$ is more
informative than $\mu$ if and only if $\mu$ can be replicated by
chaining the output of $\kappa$ through an additional channel
$\lambda$. Next, we present a rigorous formulation of these
definitions and results in terms of random variables.

Consider three random variable $X, Y, Z$ with finite state spaces
$\Xcal, \Ycal, \Zcal$.  Suppose that an agent has a finite set of
possible actions~$\Acal$.  After the agent chooses her
action~$a\in\Acal$, she receives a reward $u(x,a)$, which not only
depends on the chosen action $a\in\Acal$, but also on the value
$x\in\Xcal$ of the random variable~$X$.  The tuple $(p,\Acal,u)$,
consisting of the prior distribution~$p$ of $X$, the set of possible
actions~$\Acal$ and the reward function~$u$ is called a \emph{decision
  problem}.  If the agent can observe the value $x$ of $X$ before
choosing her action, her best strategy is to chose $a$ such that
$u(x,a) = \max_{a'\in\Acal}u(x,a')$.

Suppose that the agent cannot observe $X$ directly, but the agent knows the probability distribution $p$ of~$X$.
Moreover, the agent observes a random variable $Y$ with conditional distribution equal to~$\kappa(x;Y)$, where $\kappa$
belongs to the set
\begin{equation*}
  K(\Xcal;\Ycal) = \Big\{ \kappa\in[0,1]^{\Xcal\times\Ycal} \,:\, \sum_{y\in\Ycal}\kappa(x;y) = 1\text{ for all }x \Big\} 
\end{equation*}
of (row) stochastic matrices.  $\kappa$ will also be called a \emph{channel} from $\Xcal$ to~$\Ycal$, and $\Xcal$ is
the \emph{domain} 
of~$\kappa$.  When using a channel~$\kappa$, the agent's optimal
strategy is to choose her action such that her expected reward
\begin{equation}
  \label{eq:exp-reward}
  \sum_{x}P(X=x|Y=y)u(x,a) = \frac{\sum_{x}p(x)\kappa(x;y) u(x,a)}{\sum_{x\in\Xcal}p(x)\kappa(x;y)}
\end{equation}
is maximal.  Note that, in order to maximize~\eqref{eq:exp-reward}, the agent has to know (or estimate) the prior
distribution of~$X$ as well as the channel~$\kappa$.  Often, the agent is allowed to play a stochastic strategy.
However, in the present setting, the agent cannot increase her expected reward by randomizing her actions, and
therefore, we only consider deterministic strategies here.  Probabilistic strategies do sometimes lead to a nicer
mathematical formulation, see Remark~\ref{rem:prob-strat} below.

Let $R(\kappa,p,u,y)$ be the maximum of~\eqref{eq:exp-reward} (over $a\in\Acal$), and let
\begin{equation*}
  R(\kappa,p,u) = \sum_{x}P(Y=y) R(\kappa,p,u,y)
\end{equation*}
be the maximal expected reward that the agent can achieve by always choosing the optimal action.
In this setting, we make the following definition:
\begin{defn}
  \label{def:unq-info}
  Let $X,Y,Z$ be three random variables, let $p$ be the marginal distribution of~$X$, and let $\kappa\in
  K(\Xcal;\Ycal)$, $\mu\in K(\Xcal;\Zcal)$ such that
  \begin{omultline}
    \label{eq:XYkappamu}
    P(X=x,Y=y) = p(x)\kappa(x;y) \onl
    \quad\text{ and }\quad
    P(X=x,Z=z) = p(x)\mu(x;z).
  \end{omultline}
  $Y$ is \emph{more informative} about $X$ than~$Z$, if for any
  decision problem $(p,\Acal,u)$ the inequality $R(\kappa,p,u)\ge
  R(\mu,p,u)$ holds.  In this situation we also say that $Y$ knows
  everything that $Z$ knows about~$X$, and we write $Y\uge_{X} Z$.
\end{defn}
More generally, one could try to quantify how much of the information
that $Y$ has about~$X$ is unknown to~$Z$, i.e. the unique information
of $Y$, as compared to how much of the information about $X$ is shared
between $Y$ and $Z$. Here, we do not further explore this idea, but
the paper~\cite{BROJA13:Quantifying_unique_information} discusses one
way to decompose the mutual information $MI(X:\{Y,Z\})$ along those
lines; see also references therein for other possible decompositions.

Definition~\ref{def:unq-info} compares the random variables $Y$ and $Z$ using only the marginal distribution $p$ of~$X$
and the conditional distributions $\kappa$, $\mu$ of $Y$ and~$Z$, respectively, given~$X$.  It is also possible to
ignore the marginal distribution of $p$ and directly compare the two channels $\kappa$ and~$\mu$.  The information
of~$X$ is passed through these channels and then needs to be decoded in view of the decision problem at hand.  The
channels can be applied to arbitrary random variables (having the same state space), that is, the marginal distribution
$p$ of~$X$ may be arbitrary.  As an example, one may think about a device that measures a physical observable (possibly
with stochastic noise).  Depending on the experiment in which this device is used, the prior distribution of the
observable changes.  Nevertheless, the characteristics of the measurement error do not depend on the experiment.  In
this setting we make the following definition, which corresponds to Blackwell's original definition:
\begin{defn}
  \label{def:unq-info-channel}
  Let $\kappa\in K(\Xcal;\Ycal)$ and $\mu\in K(\Xcal;\Zcal)$ be two channels with domain~$\Xcal$.
  $\kappa$ is \emph{more informative} than~$\mu$, if for any decision problem $(p,\Acal,u)$ the
  inequality $R(\kappa,p,u)\ge R(\mu,p,u)$ holds.  In this situation we also write $\kappa\uge_{\Xcal}\mu$.
\end{defn}

Definitions~\ref{def:unq-info} and~\ref{def:unq-info-channel} define two relations $\uge_{X}$ and
$\uge_{\Xcal}$ on the sets of random variables and on the set of stochastic matrices with domain~$\Xcal$.  These
two relations are preorders, that is, they are reflexive and transitive.  They are related to the following preorders:
\begin{defn}
  \label{def:majorization-order}
  \begin{itemize}
  \item Write $Y\mge_{X} Z$ if there is a random variable~$Z'$ such that the following two conditions hold:
    \begin{enumerate}
    \item The Markov chain $X\to Y\to Z'$ holds (that is, $X$ and $Z'$ are conditionally independent given~$Y$).
    \item The pairs $(X,Z)$ and $(X,Z')$ have the same distribution.
    \end{enumerate}
  \item Write $\kappa\mge_{\Xcal}\mu$ if there is a stochastic matrix $\lambda$ such that~$\mu=\kappa\lambda$.
  \end{itemize}
\end{defn}
The order $\mge_{\Xcal}$ is sometimes also called the \emph{majorization order}~\cite{MarshallOlkin79:Inequalities}.

Clearly, if there is a Markov chain $X\xrightarrow{\kappa} Y\xrightarrow{\lambda} Z$, then $Y$ knows everything that $Z$
about~$X$,
and the same is true if there is a Markov chain $X\xrightarrow{\kappa} Y\xrightarrow{\lambda} Z'$ for a variable $Z'$ that has the same joint distribution with~$X$ as~$Z$.
Similarly, if $\mu=\kappa\lambda$, then $\kappa\uge_{\Xcal}\mu$.  Therefore, $Y\mge_{X} Z$
implies $Y\uge_{X}Z$, and $\kappa\mge_{\Xcal}\mu$ implies $\kappa\uge_{\Xcal}\mu$.  Surprisingly, the
converse implications also hold, as the following result shows:
\begin{thm}[Sherman-Stein-Blackwell theorem]$ $
  \label{thm:Blackwells-thm-0}
  \begin{enumerate}
  \item $\kappa\mge_{\Xcal}\mu$ if and only if $\kappa\uge_{\Xcal}\mu$.
  \item $Y\mge_{X}Z$ if and only if $Y\uge_{X}Z$.
  \item Let $X,Y,Z$ be three random variables, let $p$ be the marginal distribution of~$X$.  Assume that $p(x)>0$ for
    all~$x\in\Xcal$, and let $\kappa\in K(\Xcal;\Ycal)$, $\mu\in K(\Xcal;\Zcal)$ be the conditional distributions of $Y$ and $Z$ given~$X$.
    Then $\kappa\uge_{\Xcal}\mu$ if and only if $Y\uge_{\Xcal}Z$.
  \end{enumerate}
\end{thm}
\begin{oproof}
  See Section~3 in~\cite{Blackwell53:Equivalent_comparisons_of_experiments} for a short proof of~1).
  To prove~3), observe that $\kappa\uge_{\Xcal}\mu$ implies $Y\uge_{\Xcal}Z$ by definition.  For the other direction,
  let $u$ be a reward function and let $q$ be a probability distribution for~$X$.  Define a reward function $u'$ via
  $u'(x,a)=\frac{q(x)}{p(x)}u(x,a)$.  Then
  \begin{equation*}
    \frac{\sum_{x}p(x)\kappa(x;y) u'(x,a)}{\sum_{x\in\Xcal}p(x)\kappa(x;y)}
    = \frac{\sum_{x}q(x)\kappa(x;y) u(x,a)}{\sum_{x\in\Xcal}p(x)\kappa(x;y)},
  \end{equation*}
  and thus the optimal strategy when $X$ is distributed according to $q$ and the reward is~$u$ is the same as the
  optimal strategy when $X$ is distributed according to $p$ and the reward is~$u'$.  Therefore,
  $R(\kappa,q,u)=R(\kappa,p,u')\ge R(\mu,p,u')=R(\mu,q,u)$.  Finally, 2)~is a direct consequence of~1)~and~3).
\end{oproof}
We call Theorem~\ref{thm:Blackwells-thm-0} Blackwell's theorem, since the most difficult part of the proof is the
equivalence of 1.~and~3., which is usually called Blackwell's theorem.  However, in the following we only use the case
that all state spaces $\Xcal,\Ycal$ and $\Zcal$ are finite, and in this case the result was proven by Sherman and Stein;
see references in~\cite{Blackwell53:Equivalent_comparisons_of_experiments}.
In the following, we call $\uge_{\Xcal}$ and $\uge_{X}$ the \emph{Blackwell preorders}.

The rest of this manuscript is organized as follows: In Section~\ref{sec:zonotope-order} we define the zonotope
preorder, which is weaker than the Blackwell preorder.  To show that the Blackwell order does not define a lattice,
it is sufficient to show that the zonotope preorder does not define a lattice.  In Section~\ref{sec:k-dps} we mention
Blackwell's $k$-decision orders, which generalize the zonotope order.  In Section~\ref{sec:binary-case} we discuss the
binary case, i.e. $|\Xcal| = 2$. In this special case, all $k$-decision orders agree and define a lattice.  In Section~\ref{sec:ge2} we
show that this is not true for~$|\Xcal|>2$.

Most of the mathematical results of this contribution are known; except a slight generalization of Blackwell's theorem
to random variables.  However, the fact that Blackwell's relation does not define a lattice seems to be widely unknown.

\section{The zonotope order}
\label{sec:zonotope-order}

For any finite sets $\Xcal,\Ycal$ denote by $K(\Xcal;\Ycal)$ the set of (row) stochastic matrices, and denote by
$K(\Xcal)$ the union of all sets $K(\Xcal;\Ycal)$ for arbitrary finite sets~$\Ycal$.\footnote{$K(\Xcal)$ may not be a
  set in the strict set theoretical sense.  Later, only the cardinality of the target sets~$\Ycal$ plays a role, and if
  we identify sets of the same (finite) cardinality and the corresponding stochastic matrices, we obtain again a set.}
In this section we define another preorder on $K(\Xcal)$.  To each stochastic matrix $\kappa\in K(\Xcal;\Ycal)$ and
$y\in\Ycal$ denote by $\kappa_{y}$ the $y$th column of~$\kappa$.  The zonotope of $\kappa$ is the convex set
\begin{omultline*}
  Z_{\kappa} = \left\{ \sum_{y\in\Ycal}a_{y}\kappa_{y} : a_{y}\in[0;1]\text{ for all }y\in\Ycal\right\}
  \\
  = \Big\{ \kappa a : a\in[0;1]^{\Ycal}\Big\}.
\end{omultline*}
In other words, $Z_{\kappa}$ is the image of the unit cube $[0,1]^{\Ycal}$ under the linear map corresponding
to~$\kappa$.  The zonotope $Z_{\kappa}$ is a polytope.  Every vertex of $Z_{\kappa}$ is an image of a vertex of the
hypercube $[0;1]^{\Ycal}$ under~$\kappa$, that is, the vertices of $Z_{\kappa}$ are a subset
of~$\kappa(\{0,1\}^{\Ycal})$.

Denote by $\zeros_{\Xcal}$ and $\ones_{\Xcal}$ the vectors $(0,\dots,0),(1,\dots,1)\in\Rb^{\Xcal}$.  The zonotope of
$\kappa$ has the following properties:
\begin{lemma}
  \label{lem:elem-props-Zkappa}
  \begin{itemize}
  \item $Z_{\kappa}\subseteq [0,1]^{\Xcal}$.
  \item $Z_{\kappa}$ contains the vertices $\zeros_{\Xcal}$ and~$\ones_{\Xcal}$.  Hence the diagonal
    $\Dia=\{(r,r,r):0\le r\le1\}$ of the hypercube $[0,1]^{\Xcal}$ is a subset of $Z_{\kappa}$.
  \item $Z_{\kappa}$ has the following symmetry: If $v\in Z_{\kappa}$, then $\ones_{\Xcal}-v\in Z_{\kappa}$.  In fact,
    if~$v = \kappa a$, then $\ones_{\Xcal}-v = \kappa(\ones_{\Xcal} - a)$.
  \end{itemize}
\end{lemma}
\begin{defn}
  \label{def:zonotope-order}
  The zonotope order is the preorder $\zge_{\Xcal}$ on $K(\Xcal)$ defined as follows:
  For $\kappa,\mu\in K(\Xcal)$ the relation $\kappa\zge_{\Xcal}\mu$ holds if and only if $Z_{\kappa}\supseteq Z_{\mu}$.
\end{defn}
The zonotope order is related to the Blackwell order:
\begin{lemma}
  \label{lem:zonotopes}
  Let $\kappa,\mu\in K(\Xcal)$.  If $\kappa\mge_{\Xcal}\mu$, then $\kappa\zge_{\Xcal}\mu$.
\end{lemma}
\begin{oproof}
  Observe that~$Z_{\mu}=(\kappa\lambda)([0,1]^{\Zcal}) = \kappa(Z_{\lambda})\subseteq\kappa([0,1]^{\Ycal})=Z_{\kappa}$.
\end{oproof}
In general, the converse of Lemma~\ref{lem:zonotopes} is false, unless $|\Xcal|=2$, see Sections~\ref{sec:binary-case}
and~\ref{sec:ge2}.  The following weaker statement always holds:
\begin{lemma}
  \label{lem:zonotope-equality}
  Let $\kappa,\mu\in K(\Xcal)$.  If $Z_{\mu}=Z_{\kappa}$, then $\mu\mge_{\Xcal}\kappa$ and $\kappa\mge_{\Xcal}\mu$.
\end{lemma}
\begin{oproof}
  Each zonotope has a unique minimal set of generators, which consists of the set of edge vectors.  Let
  $\nu_{1},\dots,\nu_{r}$ be the minimal generators of the zonotope~$Z_{\kappa} = Z_{\mu}$.  The vectors $\nu_{1},\dots,\nu_{r}$ form
  the columns of a stochastic matrix $\nu$ (indeed, they are non-negative, and their sum equals~$\ones_{\Xcal}$).  It
  suffices to show that $\mu\mge_{\Xcal}\nu$ and $\nu\mge_{\Xcal}\mu$.

  The rows $\mu_{y}$, $y\in\Ycal$, of $\mu$ are another set of generators of~$Z_{\mu}$.  If a generating set of a
  zonotope is not minimal, then some of these generators must be proportional, and every generator is proportional to a
  minimal generator.  More precisely, there is a partition $\Ycal=\Ycal_{1}\cup\dots\cup\Ycal_{r}$ of $\Ycal$ and there
  are scalars $a_{y}$ for all $y\in\Ycal$ such that the following holds:
  \begin{itemize}
  \item $\nu_{i}=\sum_{y\in\Ycal_{i}}\mu_{y}$ for all $i=1,\dots,r$.
  \item $\mu_{y}=a_{y}\nu_{i}$ for all $i=1,\dots,r$ and for all $y\in\Ycal_{i}$.
  \end{itemize}
  Observe that $\nu_{i}=\sum_{y\in\Ycal_{i}}a_{y}\nu_{i}$, and thus $\sum_{y\in\Ycal_{i}}a_{y}=1$.  Let
  $\lambda^{(1)}\in\Rb^{\Ycal\times r}$, $\lambda^{(2)}\in\Rb^{r\times \Ycal}$ be the matrices with matrix elements
  \ifIEEE
  \begin{align*}
    (\lambda^{(1)})_{y,i} &=
    \begin{cases}
      1, & \text{ if }y\in\Ycal_{i},\\
      0, & \text{ else},
    \end{cases}, \\
    \intertext{ and }
    (\lambda^{(2)})_{i,y} &=
    \begin{cases}
      a_{y}, & \text{ if }y\in\Ycal_{i},\\
      0, & \text{ else}.
    \end{cases}
  \end{align*}
  \else
  \begin{equation*}
    (\lambda^{(1)})_{y,i} =
    \begin{cases}
      1, & \text{ if }y\in\Ycal_{i},\\
      0, & \text{ else},
    \end{cases},
    \text{ and }\quad
    (\lambda^{(2)})_{i,y} =
    \begin{cases}
      a_{y}, & \text{ if }y\in\Ycal_{i},\\
      0, & \text{ else}.
    \end{cases}
  \end{equation*}
  \fi
  $\lambda^{(1)}$ is stochastic, since every row contains precisely one non-zero entry, which is equal to one.
  $\lambda^{(2)}$ is stochastic, since $\sum_{y\in\Ycal}\lambda^{(2)}_{i,y}=\sum_{y\in\Ycal_{i}}a_{y}=1$.  Then
  $\nu=\mu\lambda^{(1)}$ and $\mu=\nu\lambda^{(2)}$.  Therefore, $\nu\mge_{\Xcal}\mu\mge_{\Xcal}\nu$.  By symmetry,
  $\nu\mge_{\Xcal}\kappa\mge_{\Xcal}\nu$, and so the statement follows.
\end{oproof}
Lemma~\ref{lem:zonotope-equality} shows that the two preorders $\uge_{\Xcal}$ and $\zge_{\Xcal}$ define the same
equivalence relation:
\begin{equation*}
  \kappa\sim\mu
  \;:\Longleftrightarrow\;
  (\kappa\uge_{\Xcal}\mu\uge_{\Xcal}\kappa)
  \;\Longleftrightarrow\;
  (\kappa\zge_{\Xcal}\mu\zge_{\Xcal}\kappa)\,.
\end{equation*}
Therefore, the two preorders $\uge_{\Xcal}$ and $\zge_{\Xcal}$ induce orders on the set~$K(\Xcal)/\sim$ of stochastic matrices
modulo the equivalence~$\sim$.  Lemma~\ref{lem:zonotopes} shows that $\zge_{\Xcal}$ is a refinement of~$\uge_{\Xcal}$.
The preorder $\zge_{\Xcal}$ can also be given an interpretation in terms of binary decision problems, that is, decision
problems where the agent has only two options:
\begin{prop}
  \label{prop:zonotope-order-binary-games}
  The following statements are equivalent:
  \begin{enumerate}
  \item
    $R(\kappa,p,u)\ge R(\mu,p,u)$ for all binary decision problems~$(p,\{0,1\},u)$.
  \item
    There exists a prior distribution~$p$ of~$X$ with full support such that $R(\kappa,p,u)\ge R(\mu,p,u)$ for all
    binary decision problems~$(p,\{0,1\},u)$.
  \item
    $\kappa\zge_{\Xcal}\mu$.
  \end{enumerate}
\end{prop}
The proof of Proposition~\ref{prop:zonotope-order-binary-games} makes use of the following Lemma, which will also be
needed later:
\begin{lemma}
  \label{lem:binary-subzonotopes}
  If $v\in Z_{\kappa}$, then the kernel $\mu_{(v)}=(v,\ones_{\Xcal}-v)$ satisfies $\kappa\mge_{\Xcal}\mu_{(v)}$ and $v\in
  Z_{\mu_{(v)}}$.
\end{lemma}
\begin{oproof}
  Since $v\in Z_{\kappa}$, there exists $a\in[0;1]^{\Ycal}$ with~$v=\kappa(a)$.  Then $\lambda=(a,\ones_{\Ycal}-a)$ is a
  stochastic matrix satisfying $\mu_{(v)}=\kappa\lambda$.
\end{oproof}
\begin{oproof}[Proof of Proposition~\ref{prop:zonotope-order-binary-games}]
  The equivalence between 1)~and 2)~can be proven as in the proof of Theorem~\ref{thm:Blackwells-thm-0}.

  Suppose there exists $v\in Z_{\mu}$ with $v\notin Z_{\kappa}$.  Then $v\in Z_{\mu_{v}}\setminus Z_{\kappa}$, and by
  Blackwell's theorem, there exist $\Acal'$, $p$ and $u'\in\Rb^{\Xcal\times\Acal'}$ such that
  $R(\mu_{(v)},p,u')>R(\kappa,p,u')$.  Consider the optimal strategy of an agent who only observes the outcome of the
  channel~$\mu_{(v)}$.  Since $\mu_{(v)}$ is binary, this optimal strategy only makes use of a two-element subset
  of~$\Acal'$, say $\Acal\subseteq\Acal'$.  Let $u$ be the restriction of~$u'$ to $\Rb^{\Xcal\times\Acal}$.  Then
  $R(\mu,p,u)\ge R(\mu_{(v)},p,u)=R(\mu_{(v)},p,u')>R(\kappa,p,u')\ge R(\kappa,p,u)$, where the first inequality
  follows from $\mu\mge\mu_{(v)}$ (by Lemma~\ref{lem:binary-subzonotopes}), and the last inequality comes from the fact
  that an agent who observes $\kappa$ cannot perform better if its options are restricted from $\Acal'$ to~$\Acal$.
  Therefore, if $Z_{\mu}$ is not a sub-zonotope of $Z_{\kappa}$, then there is a binary decision problem in which $\mu$
  performs better than~$\kappa$.

  Conversely, assume that $\mu$ performs better than~$\kappa$ in some binary decision problem with $\Acal=\{0,1\}$ and
  reward function~$u$.  Fix some optimal strategy, let $\Zcal_{0}$ be the subset of $\Zcal$ where the agent chooses
  action~$0$, and let $\Zcal_{1}=\Zcal\setminus\Zcal_{0}$.  Let $v=\sum_{z\in\Zcal_{0}}\mu_{z}$ be the sum of the
  columns of $\mu$ indexed by~$\Zcal_{0}$.  Then $\mu\mge\mu_{v}$, and so $R(\mu,p,u)\ge R(\mu_{v},p,u)$.  In fact,
  since the agent only needs to know which action it has to choose (it can forget any other information contained in the
  output of the channel~$\mu$), it follows that~$R(\mu_{v},p,u)=R(\mu,p,u)>R(\kappa,p,u)$.  By
  Lemma~\ref{lem:binary-subzonotopes} and Blackwell's theorem, $Z_{\mu_{v}}$ is not a sub-zonotope of~$Z_{\kappa}$.
  Therefore, $Z_{\mu}$ (which contains $Z_{\mu_{v}}$) is not a sub-zonotope of~$Z_{\kappa}$ either.
\end{oproof}
\begin{rem}
  \label{rem:prob-strat}
  The proof of Proposition~\ref{prop:zonotope-order-binary-games} implicitly uses the following construction: The action
  that the agent chooses can be considered as a random variable~$A$.  This is true both in the case that the agent has a
  deterministic strategy (depending on the observed random variable~$Y$) and in the case that the agent is allowed to
  have a probabilistic strategy.  $A$ satisfies the Markov chain $X\to Y\to A$.  Denote by $\mu_{A}\in K(\Xcal;\Acal)$
  the stochastic matrix describing the probability of the agent's action given~$X$.  Then $\kappa\mge_{\Xcal}\mu_{A}$.
  Conversely, any $\mu\in K(\Xcal)$ satisfying $\kappa\mge_{\Xcal}\mu$ can be interpreted as a probabilistic strategy
  that the agent may use.

  By Lemma~\ref{lem:binary-subzonotopes}, any $v\in Z_{\kappa}$ corresponds to a kernel $\mu_{(v)}$ lying below~$\kappa$
  in the Blackwell order.  Such a kernel can be interpreted as a stochastic binary strategy.  Therefore, the elements
  of~$Z_{\kappa}$ correspond to the set of stochastic binary strategies that the agent who observes $Y$ may play.
\end{rem}

\begin{rem}
  \label{rem:Lorenz-zonoids}
  The zonotope $Z_{\kappa}$ is related to the Lorenz zonoid that is used to quantify the disparity of the joint
  distribution of goods in a population~\cite{KoshevoyMosler96:Lorenz_zonoid}.  In the case of a finite population, the
  zonoid becomes a zonotope.  The inclusion of Lorenz zonoids defines an order, called \emph{Lorenz zonoid order}.
\end{rem}

\section[k-decision problems]{$k$-decision problems.}
\label{sec:k-dps}

In
~\cite{Blackwell53:Equivalent_comparisons_of_experiments}, Blackwell also introduces the following preorders:
\begin{defn}
  For any $k>1$, the $k$-decision order is the preorder defined on $K(\Xcal)$ as follows: For $\kappa,\mu\in K(\Xcal)$
  the relation $\kappa\zge_{\Xcal;k}\mu$ holds if and only if for any decision problem $(p,\Acal,u)$ with $|\Acal|\le k$
  the inequality $R(\kappa,p,u)\ge R(\mu,p,u)$ holds.
\end{defn}

\begin{lemma}
  \begin{enumerate}
  \item $\kappa\zge_{\Xcal;2}\mu$ if and only if $\kappa\zge_{\Xcal}\mu$.
  \item $\kappa\zge_{\Xcal;k+1}\mu$ implies $\kappa\zge_{\Xcal;k}\mu$ for all~$k\ge 2$.
  \item $\kappa\uge_{\Xcal}\mu$ implies $\kappa\zge_{\Xcal;k}\mu$ for all~$k\ge 2$.
  \item If $\kappa\zge_{\Xcal;k}\mu$ for all~$k\ge 2$, then $\kappa\uge_{\Xcal}\mu$.
  \item If $\kappa\in K(\Xcal;\Ycal)$, $\mu\in K(\Xcal;\Zcal)$ and $\kappa\zge_{\Xcal;k}\mu$ with $k\ge|\Zcal|$, then
    $\kappa\uge_{\Xcal}\mu$.
  \end{enumerate}
\end{lemma}
\begin{oproof}
  Statement~1) follows from Proposition~\ref{prop:zonotope-order-binary-games}.  Statements~2) to~4) are direct.
  Statement~5) follows since a deterministic strategy cannot make use of more actions than the number of possible
  outputs.
\end{oproof}

According to an unpublished paper of Stein (cited in~\cite{Blackwell53:Equivalent_comparisons_of_experiments}), the
preorders $\zge_{\Xcal;2},\zge_{\Xcal;3},\dots,\zge_{\Xcal;|\Xcal|}$ are all different in general.  In
Section~\ref{sec:ge2}, we give an example that shows that $\zge_{\Xcal;2}=\zge_{\Xcal}$ is different from~$\uge_{\Xcal}$
and from~$\zge_{\Xcal;3}$ for~$|\Xcal|=3$.

\section[The binary case]{The case $|\Xcal|=2$.}
\label{sec:binary-case}

In this section we collect some aspects in which the binary case $|\Xcal|=2$ is different from the general case.  First,
the zonotope order and the Blackwell order are identical (and hence, all the $k$-decision orders agree).  Moreover, the
Blackwell order defines a lattice.  These results are well-known.  We derive them
from~\cite{Blackwell53:Equivalent_comparisons_of_experiments}; see~\cite{Dahl98:Matrix_majorization} for a proof from
the perspective of majorization theory.
\begin{prop}
  \label{prop:binary-zon-is-Black}
  Let $\kappa,\mu\in K(\{0,1\})$.  Then $\kappa\zge_{\{0,1\}}\mu$ if and only if $\kappa\mge_{\{0,1\}}\mu$.
\end{prop}
\begin{oproof}
  See~\cite[Theorem~10]{Blackwell53:Equivalent_comparisons_of_experiments}.
\end{oproof}
\begin{prop}
  \label{prop:binary-lattice}
  If~$|\Xcal|=2$, then the Blackwell order defines a lattice.
\end{prop}
The proof makes use of the following lemma:
\begin{lemma}
  \label{lem:bin-zon}
  If $|\Xcal|=2$, then a zonotope is of the form $Z=Z_{\kappa}$ for some $\kappa\in K(\Xcal)$ if and only if $Z$
  satisfies the properties of Lemma~\ref{lem:elem-props-Zkappa}.
\end{lemma}
\begin{oproof}
  Due to symmetry, the number of vertices of $Z$ is even; say~$2n$.  Let $x_{0}=\zeros_{\Xcal},x_{1},\dots,x_{2n-1}$ be
  the vertices of $Z$ ordered counterclockwise.  Then~$x_{n}=\ones_{\Xcal}$.  Let $\kappa_{i}=x_{i}-x_{i-1}$ for
  $i=1,\dots,n$ be the edge vectors of~$Z$, and let $\kappa\in\Rb^{\Xcal\times n}$ be the matrix with columns
  $\kappa_{1},\dots,\kappa_{n}$.  By convexity of~$Z$, all vectors $\kappa_{1},\dots,\kappa_{n}$ are non-negative.
  Moreover, $\kappa_{1}+\dots+\kappa_{n}=\ones_{\Xcal}$.  Hence, $\kappa$ is a stochastic matrix.  Clearly,
  $Z=Z_{\kappa}$.
\end{oproof}
\begin{oproof}[Proof of Proposition~\ref{prop:binary-lattice}]
  It suffices to show that the zonotope order defines a lattice.  For $\kappa,\mu\in K(\Xcal)$ let
  $Z_{\wedge}=Z_{\kappa}\cap Z_{\mu}$, and let $Z_{\vee}$ be the convex hull of $Z_{\kappa}\cup Z_{\mu}$.  By
  Lemma~\ref{lem:bin-zon} there exist $\kappa_{\wedge},\kappa_{\vee}\in K(\Xcal)$ with $Z_{\wedge}=Z_{\mu_{\wedge}}$
  and~$Z_{\vee}=Z_{\mu_{\vee}}$.  Any common lower bound $\kappa'$ of $\kappa$ and $\mu$ satisfies $Z_{\kappa'}\subseteq
  Z_{\wedge}$, and any upper bound $\mu'$ satisfies $Z_{\vee}\subseteq Z_{\mu'}$.  The statement now follows from
  Proposition~\ref{prop:binary-zon-is-Black}.
\end{oproof}

\section[The general case]{The case $|\Xcal|>2$.}
\label{sec:ge2}

In this section, assume that $|\Xcal|>2$.  In this case, the converse of Lemma~\ref{lem:zonotopes} does not hold,
i.e.~there are stochastic matrices $\kappa\in K(\Xcal;\Ycal),\mu\in K(\Xcal;\Zcal)$ with $\kappa\not\mge\mu$
and~$Z_{\mu}\subseteq Z_{\kappa}$.  This fact is mentioned in~\cite{Dahl98:Matrix_majorization}, but no example is
given.  By Lemma~\ref{lem:binary-subzonotopes}, in such an example the matrix $\mu$ must have at least three columns.
Here comes an example:

\begin{ex}
  Let
  \begin{equation*}
    \kappa=
    \begin{pmatrix}
      \frac12 & 0 & 0 & \frac12 \\
      0 & \frac12 & 0 & \frac12 \\
      0 & 0 & \frac12 & \frac12 \\
    \end{pmatrix}
    \quad\text{ and }\quad
    \mu=
    \begin{pmatrix}
      \frac12 & \frac12 & 0 \\
      \frac12 & 0 & \frac12 \\
      0 & \frac12 & \frac12 \\
    \end{pmatrix}.
  \end{equation*}
  For each $i\in\{1,2,3\}$ there is a unique way of writing $\mu_{i}$ as a linear combination of
  $\kappa_{1},\dots,\kappa_{4}$ with non-negative coefficients.  That is,
  \begin{equation*}
    \lambda=
    \begin{pmatrix}
      1 & 1 & 0 \\
      1 & 0 & 1 \\
      0 & 1 & 1 \\
      0 & 0 & 0 \\
    \end{pmatrix}
  \end{equation*}
  is the unique non-negative matrix satisfying $\mu=\kappa\lambda$.  Alas, $\lambda$ is not stochastic, and so
  $\kappa\not\mge\mu$.  However, $Z_{\mu}\subset Z_{\kappa}$.  To show this, it suffices to show that all vertices
  of~$Z_{\mu}$ lie in~$Z_{\kappa}$, or that all vectors of the form $\mu_{i}$ or $\mu_{i}+\mu_{j}$ (for $i\neq j$) lie
  in~$Z_{\mu}$ (the vector $\mu_{1}+\mu_{2}+\mu_{3}=\ones_{\Xcal}$ lies in~$Z_{\mu}$).  Since all coefficients of
  $\lambda$ lie in~$[0,1]$, the vectors $\mu_{i}$ lie in~$Z_{\kappa}$, and hence $\ones_{\Xcal}-\mu_{i}\in Z_{\kappa}$.
  Moreover, the relation $\mu_{1}+\mu_{2}+\mu_{3}=\ones_{\Xcal}$ shows that $\mu_{i}+\mu_{j}=\ones_{\Xcal}-\mu_{k}$ for
  any permutation $(i,j,k)$ of $(1,2,3)$.  Therefore, all vertices of $Z_{\mu}$ lie in~$Z_{\kappa}$.

  It is also possible to find a concrete decision problem for which $\mu$ is better than~$\kappa$: Let
  \begin{equation*}
    u =
    \begin{pmatrix}
      -5 & 1 & 1 \\
      1 & -5 & 1 \\
      1 & 1 & -5
    \end{pmatrix}.
  \end{equation*}
  In this case, there are three different actions.  Action $i$ gives a reward of $+1$ if $X\neq i$ and a penalty of $-5$
  if~$X=i$.  Using the channel $\mu$ it is possible to obtain the expected reward~$+1$, while the maximal expected
  reward when using channel~$\kappa$ is $-2$.  This shows that the preorders $\zge_{\Xcal}$ and $\zge_{\Xcal;2}$ are
  indeed different.
\end{ex}

\begin{thm}
  \label{thm:no-lattice}
  If $|\Xcal|>2$, then the preorders 
  $\uge$ and $\zge$ do not define lattices on~$K(\Xcal)$.  In general, neither greatest lower bounds nor least upper
  bounds exist.
\end{thm}

\begin{oproof}
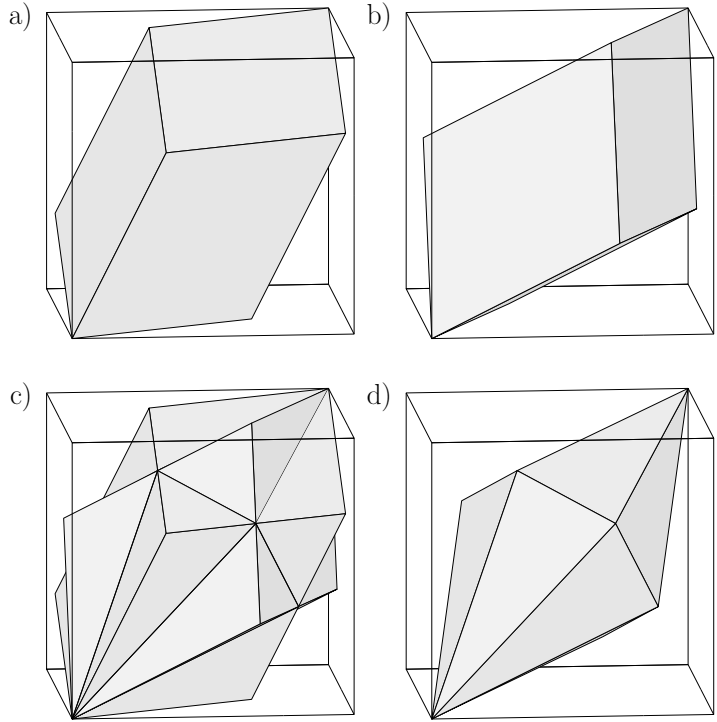
\begin{figure}
  \centering \resizebox{\ifIEEE\linewidth\else 0.6\textwidth\fi}{!}{
\begin{tikzpicture}[line join=round]
\draw(7.233,-2.381)--(10.173,-2.333)--(10.44,-2.857);
\draw(10.173,3.476)--(4.293,3.381)--(4.293,-2.429)--(10.173,-2.333)--(10.173,3.476);
\draw(-.267,-10.381)--(2.673,-10.333)--(2.94,-10.857);
\draw(2.673,-4.524)--(-3.207,-4.619)--(-3.207,-10.429)--(2.673,-10.333)--(2.673,-4.524);
\draw(7.234,-10.381)--(10.173,-10.333)--(10.44,-10.857);
\draw(10.173,-4.524)--(4.293,-4.619)--(4.293,-10.429)--(10.173,-10.333)--(10.173,-4.524);
\draw(2.673,3.476)--(-3.207,3.381)--(-3.207,-2.429)--(2.673,-2.333)--(2.673,3.476);
\draw(-2.673,-3.476)--(-3.207,-2.429)--(2.673,-2.333)--(2.94,-2.857);
\draw(4.827,-11.476)--(4.293,-10.429)--(7.234,-10.381);
\draw(-2.673,-11.476)--(-3.207,-10.429)--(-.267,-10.381);
\draw(4.827,-3.476)--(4.293,-2.429)--(7.233,-2.381);
\filldraw[fill=gray!15](2.673,3.476)--(-1.069,3.063)--(-.713,.429)--(3.029,.841)--cycle;
\draw(2.673,3.476)--(3.207,2.429)--(.267,2.381);
\filldraw[fill=gray!25](9.55,-9.112)--(10.173,-4.524)--(8.659,-7.364)--cycle;
\filldraw[fill=gray!15](8.659,-7.364)--(10.173,-4.524)--(6.608,-6.253)--cycle;
\draw(10.173,-4.524)--(10.707,-5.571)--(7.769,-5.619);
\draw(2.673,-4.524)--(-2.049,-6.887);
\draw(-2.049,-6.887)--(2.673,-4.524)--(-1.159,-8.636)--(-2.05,-6.889);
\draw(-1.159,-8.636)--(2.673,-4.524)--(.892,-9.747)--(-1.159,-8.636);
\draw(.892,-9.747)--(2.673,-4.524);
\filldraw[fill=gray!25,draw=none](2.673,-4.524)--(1.158,-7.365)--(2.049,-9.111)--(2.851,-8.746)--cycle;
\draw(2.049,-9.111)--(2.851,-8.746)--(2.673,-4.524);
\filldraw[fill=gray!15,draw=none](2.673,-4.524)--(-1.069,-4.937)--(-.891,-6.254)--(1.158,-7.365)--cycle;
\draw(2.673,-4.524)--(-1.069,-4.937)--(-.891,-6.254);
\draw(-.89,-6.253)--(2.673,-4.524);
\draw(1.159,-7.364)--(2.673,-4.524);
\draw(2.673,-4.524)--(-.89,-6.253);
\filldraw[fill=gray!25,draw=none](2.673,-4.524)--(1.069,-5.254)--(1.158,-7.365)--cycle;
\draw(2.673,-4.524)--(1.069,-5.254)--(1.158,-7.365);
\draw(2.673,-4.524)--(2.05,-9.11);
\draw(2.673,-4.524)--(1.159,-7.364)--(1.16,-7.366);
\draw(2.05,-9.11)--(2.673,-4.524);
\filldraw[fill=gray!15,draw=none](2.673,-4.524)--(1.158,-7.365)--(3.029,-7.159)--cycle;
\draw(1.158,-7.365)--(3.029,-7.159)--(2.673,-4.524);
\draw(2.673,-4.524)--(3.074,-5.31);
\filldraw[fill=gray!25](10.173,3.476)--(8.569,2.746)--(8.747,-1.476)--(10.351,-.746)--cycle;
\draw(10.173,3.476)--(10.574,2.69);
\draw(-2.94,2.857)--(-3.207,3.381);
\draw(7.769,-5.619)--(4.827,-5.667)--(4.293,-4.619);
\draw(-2.94,-5.143)--(-3.207,-4.619);
\draw(7.767,2.381)--(4.827,2.333)--(4.293,3.381);
\draw(-2.673,-11.476)--(-1.159,-8.636)--(.891,-9.746);
\draw(-2.05,-6.889)--(-1.159,-8.636)--(-2.673,-11.476);
\filldraw[fill=gray!30,draw=none](-2.673,-11.476)--(-1.069,-10.746)--(2.851,-8.746)--(2.049,-9.111)--cycle;
\draw(-2.673,-11.476)--(-1.069,-10.746)--(2.851,-8.746)--(2.049,-9.111);
\draw(.891,-9.746)--(.892,-9.747)--(-2.673,-11.476);
\draw(2.049,-9.113)--(.892,-9.747);
\filldraw[fill=gray!30](4.827,-11.476)--(8.392,-9.747)--(9.55,-9.112)--cycle;
\draw(-2.673,-11.476)--(.892,-9.747)--(2.049,-9.113);
\filldraw[fill=gray!30](4.827,-3.476)--(6.431,-2.746)--(10.351,-.746)--(8.747,-1.476)--cycle;
\filldraw[fill=gray!10,draw=none](-2.673,-11.476)--(1.158,-7.365)--(-.891,-6.254)--cycle;
\filldraw[fill=gray!20](-2.673,-11.476)--(-.713,-7.571)--(-1.069,-4.937)--(-3.029,-8.841)--cycle;
\filldraw[fill=gray!20](-2.673,-3.476)--(-.713,.429)--(-1.069,3.063)--(-3.029,-.841)--cycle;
\draw(2.94,-2.857)--(3.207,-3.381);
\draw(2.94,-10.857)--(3.207,-11.381);
\draw(10.44,-2.857)--(10.707,-3.381);
\draw(10.44,-10.857)--(10.707,-11.381);
\draw(-2.673,-11.476)--(-2.05,-6.888)--(-2.05,-6.889);
\draw(-2.05,-6.889)--(-2.05,-6.888);
\draw(-2.05,-6.888)--(-2.049,-6.887);
\draw(-2.049,-6.887)--(-2.05,-6.888)--(-.892,-6.254);
\draw(-.893,-6.254)--(-2.05,-6.888)--(-2.673,-11.476);
\filldraw[fill=gray!20](4.827,-11.476)--(6.608,-6.253)--(5.45,-6.888)--cycle;
\filldraw[fill=gray!20](-2.673,-11.476)--(1.069,-11.063)--(3.029,-7.159)--(-.713,-7.571)--cycle;
\filldraw[fill=gray!20](-2.673,-3.476)--(1.069,-3.063)--(3.029,.841)--(-.713,.429)--cycle;
\draw(2.049,-9.113)--(2.05,-9.112)--(-2.673,-11.476);
\draw(2.05,-9.112)--(2.05,-9.11);
\draw(2.05,-9.11)--(2.05,-9.112)--(2.049,-9.113);
\filldraw[fill=gray!30,draw=none](-2.673,-11.476)--(2.049,-9.111)--(1.247,-9.476)--cycle;
\draw(2.049,-9.111)--(1.247,-9.476)--(-2.673,-11.476);
\filldraw[fill=gray!25,draw=none](1.158,-7.365)--(1.247,-9.476)--(2.049,-9.111)--cycle;
\draw(1.158,-7.365)--(1.247,-9.476)--(2.049,-9.111);
\draw(2.049,-9.113)--(2.05,-9.112)--(2.049,-9.111);
\draw(2.049,-9.111)--(2.05,-9.112);
\filldraw[fill=gray!20](4.827,-11.476)--(9.55,-9.112)--(8.659,-7.364)--cycle;
\draw(-2.673,-11.476)--(2.049,-9.113);
\draw(2.049,-9.111)--(1.159,-7.364)--(1.158,-7.366);
\draw(1.16,-7.366)--(2.049,-9.111);
\draw(.267,2.381)--(-2.673,2.333)--(-2.94,2.857);
\draw(.267,-5.619)--(-2.673,-5.667)--(-2.94,-5.143);
\filldraw[fill=gray!10,draw=none](-2.673,-11.476)--(-.891,-6.254)--(-2.851,-7.254)--cycle;
\draw(-.891,-6.254)--(-2.851,-7.254)--(-2.673,-11.476);
\draw(-2.673,-11.476)--(-.892,-6.253)--(-.893,-6.254);
\draw(-.892,-6.253)--(-.89,-6.253);
\filldraw[fill=gray!10,draw=none](1.158,-7.365)--(1.069,-5.254)--(-.891,-6.254)--cycle;
\draw(1.158,-7.365)--(1.069,-5.254)--(-.891,-6.254);
\draw(-.89,-6.253)--(-.892,-6.253)--(1.158,-7.363);
\filldraw[fill=gray!10](4.827,-3.476)--(8.747,-1.476)--(8.569,2.746)--(4.649,.746)--cycle;
\filldraw[fill=gray!15,draw=none](-.891,-6.254)--(-.713,-7.571)--(1.158,-7.365)--cycle;
\draw(-.891,-6.254)--(-.713,-7.571)--(1.158,-7.365);
\draw(-.892,-6.254)--(-.892,-6.253);
\draw(-.89,-6.254)--(-.892,-6.253)--(-.892,-6.255);
\filldraw[fill=gray!10](4.827,-11.476)--(8.659,-7.364)--(6.608,-6.253)--cycle;
\draw(-.892,-6.255)--(-2.673,-11.476);
\draw(1.158,-7.363)--(-.89,-6.254);
\draw(10.574,2.69)--(10.707,2.429)--(7.767,2.381);
\draw(3.074,-5.31)--(3.207,-5.571)--(.267,-5.619);
\filldraw[fill=gray!10,draw=none](-2.673,-11.476)--(1.247,-9.476)--(1.158,-7.365)--cycle;
\draw(-2.673,-11.476)--(1.247,-9.476)--(1.158,-7.365);
\draw(1.158,-7.366)--(1.159,-7.364)--(1.158,-7.363);
\draw(1.158,-7.363)--(1.159,-7.364);
\draw(-2.673,-11.476)--(1.158,-7.366);
\draw(1.158,-7.366)--(-2.673,-11.476);
\draw(9.237,-3.405)--(10.707,-3.381)--(10.707,-.476);
\draw(-2.673,-3.476)--(3.207,-3.381)--(3.207,-.476);
\draw(9.239,-11.405)--(10.707,-11.381)--(10.707,-5.571)--(7.769,-5.619);
\draw(1.737,-11.405)--(3.207,-11.381)--(3.207,-5.571)--(.267,-5.619);
\draw(4.827,-11.476)--(9.239,-11.405);
\draw(-2.673,-11.476)--(1.737,-11.405);
\draw(4.827,-3.476)--(9.237,-3.405);
\draw(10.707,-.476)--(10.707,2.429)--(7.767,2.381);
\draw(3.207,-.476)--(3.207,2.429)--(.267,2.381);
\draw(-2.673,.881)--(-2.673,-3.476);
\draw(7.769,-5.619)--(4.827,-5.667)--(4.827,-11.476);
\draw(-2.673,-7.119)--(-2.673,-11.476);
\draw(7.767,2.381)--(4.827,2.333)--(4.827,-3.476);
\draw(.267,2.381)--(-2.673,2.333)--(-2.673,.881);
\draw(.267,-5.619)--(-2.673,-5.667)--(-2.673,-7.119);
\path (4.141,-4.71) node[left] {\Large d)};
\path (-3.359,-4.71) node[left] {\Large c)};
\path (4.141,3.29) node[left] {\Large b)};
\path (-3.359,3.29) node[left] {\Large a)};
\end{tikzpicture}
 }
  \caption{a), b) Two zonotopes. c) Their union. d) Their intersection, which happens to be not a zonotope.}
  \label{fig:zono-intersection}
\end{figure}
The proof builds on the fact, that in general, the intersection of two zonotopes is not a zonotope.  As an example,
consider the two stochastic matrices
\begin{equation*}
  \kappa_{1}=
  \begin{pmatrix}
    \frac13 & \frac23 & 0 \\
    0 & \frac13 & \frac23 \\
    \frac23 & 0 & \frac13
  \end{pmatrix}
  \quad\text{ and }\quad
  \kappa_{2}=
  \begin{pmatrix}
    \frac23 & \frac13 & 0 \\
    0 & \frac23 & \frac13 \\
    \frac13 & 0 & \frac23
  \end{pmatrix}.
\end{equation*}
Then $Z_{\kappa_{1}}\cap Z_{\kappa_{2}}$ is the convex hull of the columns of the matrix
\begin{equation*}
  \begin{pmatrix}
    0 & \frac56 & \frac26 & \frac26 & \frac16 & \frac46 & \frac46 & 1 \\
    0 & \frac26 & \frac56 & \frac26 & \frac46 & \frac16 & \frac46 & 1 \\
    0 & \frac26 & \frac26 & \frac56 & \frac46 & \frac46 & \frac16 & 1
  \end{pmatrix};
\end{equation*}
and this is not a zonotope. This can be seen from Fig.~(\ref{fig:zono-intersection}) showing $Z_{\kappa_1}$ (a),
$Z_{\kappa_2}$ (b), their union (c), and their intersetion (d), which is not a zonotope: A face of a zonotope is again a
zonotope, but $Z_{\kappa_{1}}\cap Z_{\kappa_{2}}$ contains triangular faces.

We claim that, if $Z_{\kappa_{1}}\cap Z_{\kappa_{2}}$ is not a zonotope, then $\kappa_{1}$ and $\kappa_{2}$ have more
than one greatest common lower bound with respect to $\uge_{\Xcal}$ and~$\zge_{\Xcal}$.  The argument only involves
lower bounds of the form~$\mu_{(v)}$, as defined in Lemma~\ref{lem:binary-subzonotopes}, and therefore, the argument
will be valid for both the preorder $\uge_{\Xcal}$ and the preorder~$\zge_{\Xcal}$.

If $\mu$ is a greatest common lower bound, then $Z_{\kappa_{1}}\supseteq Z_{\mu}\subseteq Z_{\kappa_{2}}$, and so
$Z_{\mu}\subseteq Z_{\kappa_{1}}\cap Z_{\kappa_{2}}$.  By Lemma~\ref{lem:binary-subzonotopes}, for any $v\in
Z_{\kappa_{1}}\cap Z_{\kappa_{2}}$ there is a kernel $\mu_{(v)}$ with $v\in Z_{\mu_{(v)}}$, $\kappa_{1}\succ\mu_{(v)}$ and
$\kappa_{2}\succ\mu_{(v)}$, and so $\mu_{(v)}$ is a lower bound for $\kappa_{1}$ and~$\kappa_{2}$.  Therefore, if $\mu$ is
the unique greatest common lower bound, by Lemma~\ref{lem:zonotopes}, $\mu$ must satisfy $v\in Z_{\mu_{(v)}}\subseteq
Z_{\mu}$ for all $v\in Z_{\kappa_{1}}\cap Z_{\kappa_{2}}$; and so $Z_{\kappa_{1}}\cap Z_{\kappa_{2}}\subseteq Z_{\mu}$
for $i=1,2$.  Therefore, $Z_{\mu}=Z_{\kappa_{1}}\cap Z_{\kappa_{2}}$, which is impossible.

Unique least upper bounds do not exist either in this case: Let $V$ be the set of vertices of $Z_{\kappa_{1}}\cap
Z_{\kappa_{2}}$.  Then both $\kappa_{1}$ and $\kappa_{2}$ are upper bounds of $\{\mu_{(v)}:v\in V\}$.  If there exists a
least upper bound $\mu$ of the finite set $\{\mu_{(v)}:v\in V\}$, then $\kappa_{i}\mge_{\Xcal}\mu\mge_{\Xcal}\mu_{(v)}$ for $i=1,2$
and $v\in V$, and hence $\bigcup_{v\in V} Z_{\mu_{(v)}}\subseteq Z_{\mu}\subseteq Z_{\kappa_{1}}\cap Z_{\kappa_{2}}$.
Hence $Z_{\mu}$ contains the convex hull of $\bigcup_{v\in V} Z_{\mu_{(v)}}$, which is equal to $Z_{\kappa_{1}}\cap
Z_{\kappa_{2}}$.  Therefore, $Z_{\mu}=Z_{\kappa_{1}}\cap Z_{\kappa_{2}}$, a contradiction.
\end{oproof}

\bibliographystyle{plain}
\bibliography{Blackwell}

\end{document}